\documentclass[10pt]{article}
\parskip=\bigskipamount
\begin{document}

{\bf Comment on ``A local hidden variable model of quantum
correlations exploiting the detection loophole''.}

Recently, N. and B. Gisin presented a local hidden variable
(l.h.v.) model exploiting the detection loophole which reproduces exactly
the quantum correlations of the singlet state, whenever the
detectors efficiency is less than or equal to 2/3 [1]. The first
aim of the present comment is to show that, modulo slight modifications,
this model also allows us to simulate the quantum correlations that
would be observed during the static realisation of a Franson-type
experiment in the case of fully efficient detectors. Such a result was
already presented in [2]. The second aim of this comment is to compare
the models developed in the references [1] and [2]. We shall show
that, beyond strong similarities between the two approaches, only the
first one makes it possible to simulate the situation in which the
correlations of the singlet state are tested with non-coplanar settings
of the polariser.

First of all, we invite the interested reader to consult at least the
references [1] and [2] before reading our comment, because we
shall systematically refer to them for technical details, in order not
to overload the presentation of this comment.

Remarkably, the procedures followed in these models present strong
similarities. In both cases, the starting point is to consider the
''linear'' model, a well-known l.h.v. model which saturates Bell's
inequalities. In this sense, this model can be considered to furnish one
of the ''best'' simulations of non-local correlations in terms of local
ones. It predicts that the correlation function between the values of
dichotomic observables measured in two different locations exhibits a
linear dependence on the difference of two arbitrary local phases.
These phases correspond to the directions of the
local polarisers observed in a Bell-like situation [1] and to phase
shifts between two arms of local interferometers in Franson-like
situations [2].  At this level, both models proceed in the following
way: 50
\% of the values of the hidden variables that contributed to this linear
dependence are erased in such a way that the correlation function now
exhibits a cosinusoidal dependence on the phase difference,
which is proportional to the quantum correlation function exhibited by
the singlet state. For sure, the price to pay is that 50 \% of
the coincident detections that appeared in the linear model have now
disappeared. In the model of [1], they are replaced by (conveniently
symmetrised between both detectors) single detections. This trick allows
the authors, by exploiting the detection loophole, to simulate a
situation for which the detectors efficiencies are equal to 2/3, in
which case one checks directly that the ratio between the rate of single
and coincident detections is equal to one, as it must be (${2\cdot
1/3\cdot 2/3\over 2/3\cdot 2/3}\ = {50 \%\over 50 \%}$).

In the case of a
Franson-like experiment with perfect detectors, the situation is
slightly different: coincident detections occur in 50 \% of the
cases, and non-coincident detections occur otherwise. The trick here is
to follow the same procedure as for the Bell situation with unefficient
detectors but, instead of associating the values of the hidden parameters
that are erased from the linear model with no-clicks in one detector, it
is sufficient to let them correspond with delayed, non-coincident,
detections in this detector. The delay time is equal to the difference
between the times of flight along the long arm and the short arm of the
interferometers present in Franson's device (see [2] for details). By
doing so, the total number of clicks is conserved, which corresponds to
the assumption of ideal detection (efficiency of the detectors equal to
one), and one gets 50 \% of non-coincident detections and 50 \% of
coincident ones, in accordance with the standard predictions in this
case. This is the essence of the model presented in [2]. Remark that a
symmetrisation procedure is now necessary in order to avoid to introduce
an artificial temporal dissymmetry between the times of appearance of
single detections and those of coincident detections. Incidentally, this
symmetrisation procedure restablishes the symmetry between both
detectors. Note that this procedure is valid only if the delayed
statistics is the same as the advanced one, which means that the phase
settings inside the interferometers used during the experiment are not
changed too quickly. This
property can be shown to be a particular case of a very general feature:
no l.h.v. model exists that could reproduce the statistics of
Franson-like experiments whenever the phase settings are changed at a
rate superior to the time delay between both arms of the
interferometers. This point is discussed with great accuracy in [2].

At first sight
one could think that the Franson-like situation corresponds to an
effective efficiency of the detectors equal to $\sqrt{0.5}$
(close to 70 \%), which is higher than 2/3, but the
effective efficiency that can be reached thanks to this procedure (when
one erases 50 \% of the statistics) is necessarily equal to 2/3
as we already showed. What differenciates both situations is the
appearance of coincident null counts  which have no counterpart in the
Franson-like situation, in which detectors are assumed to be perfect.
In this situation, 50 \% of the events are good in the sense
that their statistics violates Bell's inequalities, and 50 \% are bad,
but  bad events ALWAYS come in pairs, whereas in the situation
with unefficient detectors, non-detection events come either in single
detection station OR in both.

This was the first point of our comment: strong analogies exist
between l.h.v. models aimed at simulating the violation of Bell's
inequalities in the case of unefficient detectors and l.h.v. models that
simulate the statistics of Franson-like experiments. In summary,
deleted clicks in the former corrrespond to delayed or advanced,
non-coincident, clicks in the latter.

In fact, the general method sketched here which consists of departing
from the linear model and of ''erasing'' some values of the hidden
parameter in order to simulate the (biased) quantum correlations was
already developed earlier. For instance, in the references [3] and [4],
the authors even showed, thanks to an ingenious symmetrisation procedure,
how to simulate the statistics obtained with detectors of efficiency
equal to 77.80 \%, which lies very closely to the theoretical upper bound
on efficiencies\footnote{See [3] and [4]  for a detailed discussion
in the (realistic) case of a non-ideal visibility of the
correlation function.} (82.83 \%). At first sight, the approach followed
in [3] and [4] is the most efficient one, in comparison to the treatment
presented in  [1] (2/3 is less than 4/5). Nevertheless, the possibility
of non-coplanar settings of the polarisers in a Bell-like experiment
allows us to distinguish more finely these two approaches. In the first
model [1], the hidden variable implemented in the linear model,
essentially a direction, is isotropically distributed on the unity
sphere, a 2-dimensional surface. In the other approaches [2,3,4], the
corresponding hidden variable, essentially an angle, is distributed on
the unity circle, a 1-dimensional surface. In the last case, it is
impossible to apply the model to the case of non-coplanar settings, as
we shall now show. The correlation function of the singlet state is
cosinusoidal in the relative angle between the directions of the
settings of the polarisers. In virtue of the constraints imposed by the
requirement of locality, the hidden variable in [2,3,4], which is an
angle, must be evaluated relatively to an arbitrary direction of
reference that must be fixed before the particles leave the source. Now,
in the case of non-coplanar settings, the triangular inequality on the
sphere imposes that the relative angle between two directions is, in
general, not equal to the difference of the angles taken between these
directions and an a priori fixed direction of reference, so that the
model does not work in general. For what does concern the model developed
in [1], which does not violate the rotational invariance of the singlet
state, the extension to the case of non-coplanar settings is realised
without problem.

It is shown in [3] that one can build analytical l.h.v. models that
approach very closely the statistics of the singlet state whenever the
detectors efficiency is less than or equal to the efficiency treshhold
(82.83 \%), provided one considers coplanar settings of the polarisers
only. It is natural to ask the question whether this is still true
whenever one also considers non-coplanar settings, in which case the
amount of constraints to be fulfilled considerably increases. It could
be that l.h.v. models for which hidden variables are triplets of
orthogonal vectors of the unity sphere (or Bloch sphere) provide good
candidates for performing the job. Such a model supplies the best
presently available candidate for classical teleportation in the
sense that it minimizes the amount of classical information
necessary in order to simulate quantum teleportation (see [5] and
references therein).

\noindent Dr. Thomas Durt, post-doctoral fellow of the Fund for
Scientific Research (FWO), Flanders;

FUND, V.U.B., Pleinlaan 2, 1050,
Brussels, Belgium. e-mail: thomdurt@vub.ac.be

\hfill\break [1] N. Gisin and B. Gisin, quant-ph/9905018, A local
hidden variable model of quantum correlation exploiting the detection
loophole.
\hfill\break [2] S. Aerts, P. Kwiat, J-A. Larsson and M. Zukowski,
quant-ph/9812053, Two-photon Franson-type interference experiments are
no tests of local realism.
\hfill\break [3] J-A. Larsson,  Phys. Lett. A, 256: 245-252,
1999, Modeling the singlet state with local variables.
\hfill\break [4] E. Santos, Phys. Lett. A, 212: 10-14, 1996,
Unreliability of performed tests of Bell's inequalities using
parmetric-down converted photons.
\hfill\break [5] N-J. Cerf, N. Gisin, and S. Massar, quant-ph/99061,
Classical Teleportation of a Quantum Bit.

Acknowledgements: The author would like to thank M. Zukowski and D.
Kaszlikowski for helpful and stimulating discussions during his visit in
Gdansk in June 1999, and for financial support of the Flemish-Polish
Scientific Collaboration Program No. 007.
\end{document}